\documentclass[aps,pra,showpacs, onecolumn]{revtex4}

\usepackage{amsmath}

\begin{document}

\title{Exact temporal second-order coherence function for displaced-squeezed thermal states}

\author{Moorad Alexanian}
\email[]{alexanian@uncw.edu}

\affiliation{Department of Physics and Physical Oceanography\\
University of North Carolina Wilmington\\ Wilmington, NC
28403-5606\\}

\date{\today}

\begin{abstract}
We calculate exactly the quantum mechanical, temporal second-order coherence function for a single-mode, degenerate parametric amplifier for a system in the Gaussian state, viz., a displaced-squeezed thermal state. The calculation involves first the generation of the Gaussian state and subsequent measurements of two photons a time $\tau \geq 0$ apart. The generation of the Gaussian state by the parametric amplifier insures that the temporal second-order coherence function depends only on $\tau$.
\end{abstract}


\maketitle {}

Quantum optical systems give rise to interesting nonclassical behavior such as photon antibunching and sub-Poissonian photon statistics owing to the discreetness or photon nature of the radiation field \cite{SZ97}. These nonclassical features can best be quantified with the aid of the temporal second-order quantum mechanical correlation function $g^{(2)}(\tau)$ and experimentally studied using a Hanbury Brown--Twiss intensity interferometer modified for homodyne detection \cite{GSSRL07}.

Consider the unitary transformations
\begin{equation}
\hat{S}(-\xi)\hat{D}(-\alpha) \hat{a} \hat{D}(\alpha)\hat{S}(\xi) = \hat{a} \cosh r - \hat{a}^{\dag} \exp(i\theta) \sinh r +\alpha,
\end{equation}
and
\begin{equation}
\hat{S}(-\xi)\hat{D}(-\alpha) \hat{a}^{\dag} \hat{D}(\alpha)\hat{S}(\xi) = \hat{a}^{\dag} \cosh r - \hat{a} \exp(-i\theta) \sinh r +\alpha^*,
\end{equation}
with  the displacement $\hat{D}(\alpha)= \exp{(\alpha \hat{a}^{\dag} -\alpha^* \hat{a})}$  and the squeezing $\hat{S}(\xi)=  \exp\big{(}-\frac{\xi}{2} \hat{a}^{\dag 2} + \frac{\xi^*}{2} \hat{a}^{2} \big{ )}$ operators, where $\hat{a}$ ($\hat{a}^{\dag})$ is the photon annihilation (creation) operator, $\xi = r \exp{(i\theta)}$, and $\alpha= |\alpha|\exp{(i\varphi)}$.

The operators $(\alpha \hat{a}^{\dag} -\alpha^* \hat{a})$ and $(-\frac{\xi}{2} \hat{a}^{\dag 2} + \frac{\xi^*}{2} \hat{a}^{2})$ do not commute with their commutator and so the product $\hat{S}(-\xi)\hat{D}(-\alpha)$ cannot be expressed in the exponential form $\exp (i\hat{H}t/\hbar)$. However, we can find an operator $\hat{H}$ such that the Eqs. (1) and (2) are reproduced. The Hamiltonian for degenerate parametric amplification, in the interaction picture, is
\begin{equation}
\hat{H} = c \hat{a}^{\dag 2} + c^* \hat{a}^2 + b\hat{a} + b^* \hat{a}^\dag.
\end{equation}
One has that
\[
\exp{(i\hat{H}\tau/\hbar)}\hat{a} \exp{(-i\hat{H}\tau/\hbar)} =  \hat{a}\cosh(2|c|\tau/\hbar) -i\hat{a}^\dag \exp{(i\chi)} \sinh(2|c|\tau/\hbar) -ib^* \frac{\sinh (2|c|\tau/\hbar)}{2|c|}
\]
\begin{equation}
+ b\exp{(i\chi)} \frac{\cosh (2|c|\tau/\hbar)-1 }{2|c|},
\end{equation}
where $c=|c|\exp{(i\chi)}$. Comparing Eqs. (1) and (4) yields
\begin{equation}
\xi(\tau) = \frac{2i\tau}{\hbar}c = r(\tau) \exp{(i\theta)},
\end{equation}
where
\begin{equation}
r(\tau) = 2|c|\tau/\hbar,
\end{equation}
and
\begin{equation}
\alpha (\tau)= -ib^* \frac{\sinh r(\tau)}{2|c|}-i b\exp{(i\theta)} \frac{\cosh r(\tau)-1 }{2|c|}.
\end{equation}
Note that $r(0)=\alpha(0)=0$.

The quantum mechanical temporal second-order coherence \cite{SZ97} is given by
\begin{equation}
g^{(2)}(\tau) =\frac{\langle \hat{a}^{\dag}(t) \hat{a}^{\dag}(t+\tau) \hat{a}(t+\tau) \hat{a}(t)\rangle }{\langle \hat{a}^{\dag}(t)  \hat{a}(t)\rangle \langle \hat{a}^{\dag}(t+\tau)\hat{a}(t+\tau) \rangle},
\end{equation}
with $\tau\geq 0$, where the operators are in the interaction picture and the expectation values, representing an average over the initial photon states, are determined by the thermal density matrix
\begin{equation}
\hat{\rho_{0}} = \exp{(-\beta \hbar \omega\hat{n})}/ \textup{Tr}[\exp{(-\beta \hbar \omega \hat{n})}],
\end{equation}
where $\hat{n}= \hat{a}^{\dag} \hat{a}$, $\beta=1/(k_{B}T)$, and $\bar{n}= \textup{Tr}[\hat{\rho}_{0} \hat{n}]$. In quantum optics, one usually deals with statistically stationary fields, viz., the correlation functions of the field are invariant under time translation and so the correlation functions, for instance, the temporal second-order coherence (8), would be a function of the single variable $\tau$. Note, however, this requires that the thermal density matrix (9) commutes with the Hamiltonian (3) that governs the time development of the system. This, however, is not so. Nonetheless, we have suppressed the dependence on $t$ in the two-time correlation (8) since $g^{(2)}(\tau)$  can be expressed as
\begin{equation}
g^{(2)}(\tau) = \frac{\langle \hat{a}^{\dag}(0) \hat{a}^{\dag}(\tau) \hat{a}(\tau) \hat{a}(0)\rangle }{\langle \hat{a}^{\dag}(0)  \hat{a}(0)\rangle \langle \hat{a}^{\dag}(\tau)\hat{a}(\tau) \rangle},
\end{equation}
where all the expectation values in (10) and henceforth are traces with the Gaussian density operator, viz., a displaced-squeezed thermal state, given by
\begin{equation}
\hat{\rho}_{G}=  \hat{D}(\alpha) \hat{S}(\xi)\hat{\rho}_{0} \hat{S}(-\xi) \hat{D}(-\alpha),
\end{equation}
where $\alpha=\alpha(t)$ and $\xi = \xi(t)$.
Accordingly, the system is initially in the thermal state $\hat{\rho}_{0}$. After time $t$, the system evolves to the Gaussian state $\hat{\rho}_{G}$ and a photon is annihilated at time $t$, the system then  develops in time and after a time $\tau$ another photon is annihilated. Therefore, two photon are annihilated in a time separation $\tau$ when the system is in the Gaussian density state $\hat{\rho}_{G}$.

Now
\begin{equation}
\langle \hat{a}^{\dag}(\tau)\hat{a}(\tau) \rangle = (\bar{n}+1/2)\cosh[2(r+r(\tau))]-1/2 + |A(\tau)|^2,
\end{equation}
where
\begin{equation}
A(\tau)= \alpha\cosh r(\tau) -i \alpha^* \exp{(i\theta)} \sinh r(\tau) +\alpha(\tau),
\end{equation}
with $r(\tau)$ and $\alpha(\tau)$ defined by Eqs. (6) and (7), respectively. One has for $\tau=0$ that
\begin{equation}
\langle \hat{a}^{\dag}(0)\hat{a}(0)\rangle =(\bar{n}+1/2 )\cosh(2r) -1/2 + |\alpha|^2 .
\end{equation}

The calculation of the temporal second-order correlation function (10) gives
\begin{equation}
g^{(2)}(\tau)= 1 + \frac{n^2(\tau)+ s^2(\tau) +\big{[}\alpha A^*(\tau) +\alpha^* A(\tau)\big{]} n(\tau) -\big{[}\alpha A(\tau)\exp{(-i\theta)}+\alpha^* A^*(\tau)\exp{(i\theta)}\big{]}s(\tau)}{\langle \hat{a}^{\dag}(0)\hat{a}(0)\rangle \langle \hat{a}^{\dag}(\tau)\hat{a}(\tau) \rangle},
\end{equation}
where
\begin{equation}
n(\tau)= (\bar{n}+ 1/2)\cosh \big{(}2r+r(\tau)\big{)} -(1/2)\cosh r(\tau),
\end{equation}
and
\begin{equation}
s(\tau)= (\bar{n}+ 1/2)\sinh \big{(}2r+r(\tau)\big{)} -(1/2)\sinh r(\tau).
\end{equation}
Result (15) reduces to the known results for $\tau=0$, for the vacuum state with $\bar{n}=0$  \cite {GSSRL07} and for the thermal state with $\bar{n}\neq 0$ \cite{LDC14}, since $r(0)=0$ and $A(0)=\alpha$.

The parameters $c$ and $b$ in the degenerate parametric Hamiltonian (3) are determined by the parameters  $\alpha$ and $\xi$ of the Gaussian density of state (11) via
\begin{equation}
tc = -i\frac{\hbar}{2} r\exp(i\theta)
\end{equation}
and
\begin{equation}
tb= -i\frac{\hbar}{2}\Big{(} \alpha \exp{(-i\theta)} + \alpha^* \coth (r/2)\Big{)} r,
\end{equation}
where $t$ is the time that it takes for the system governed by the Hamiltonian (3) to generate the Gaussian density of state $\hat{\rho}_{G}$ from the initial density of state $\hat{\rho}_{0}$.

Note that the ratio $b/|c|$ from (18) and (19) depends only on the parameter $\alpha$ of the displacement operator $\hat{D}(\alpha)$ and the parameter $\xi$ of the squeezing operator  $\hat{S}(\xi)$. Hence, $\alpha(\tau)$ defined by (7) depends on $\tau$ only through $r(\tau)$, which is defined by (6). Therefore, $A(\tau)$ defined by  (13) depends on the parameters $\alpha$ and $\xi$ and on the variable $\tau$ only through $r(\tau)$ also. Accordingly, $g^{(2)}(\tau)$ depends on the parameters $\alpha$ and $\xi$ of the density $\hat{\rho}_{G}$ and on the variable $\tau$ only through $r(\tau)$. Therefore, for given $\alpha$ and $\xi$,  the temporal second-order coherence function $g^{(2)}(\tau)$ depends on $\tau$ only on $r(\tau)$ and the mean photon number $\bar{n}$ of the thermal state $\hat{\rho}_{0}$  .

\begin{newpage}
\bibliography{basename of .bib file}

\end{newpage}
\end{document}